\documentclass[aps,prl,preprint,superscriptaddress]{revtex4-1}
\usepackage{graphicx}
\usepackage{color}
\usepackage{amsfonts}
\usepackage{isomath}
\usepackage{amsmath}
\usepackage{amsthm}
\usepackage{amsfonts}
\usepackage{braket}

\usepackage{yhmath}

\def\WS2{WS$_2$}
\def\MoS2{MoS$_2$}{

\renewcommand{\vec}[1]{\mathbfit{#1}}

\begin{document}

\title{Observation of Electrically Tunable van Hove Singularities in Twisted Bilayer Graphene from nanoARPES}

\author{Alfred J. H. Jones$^{\dagger}$}
\affiliation{Department of Physics and Astronomy, Aarhus University, 8000 Aarhus C, Denmark}
\author{Ryan Muzzio$^{\dagger}$}
\affiliation{Department of Physics, Carnegie Mellon University, Pittsburgh, Pennsylvania 15213, USA}
\author{Paulina Majchrzak$^{\dagger}$}
\author{Sahar Pakdel}
\author{Davide Curcio}
\author{Klara Volckaert}
\author{Deepnarayan Biswas}
\affiliation{Department of Physics and Astronomy, Aarhus University, 8000 Aarhus C, Denmark}
\author{Jacob Gobbo}
\author{Simranjeet Singh}
\affiliation{Department of Physics, Carnegie Mellon University, Pittsburgh, Pennsylvania 15213, USA}
\author{Jeremy T. Robinson}
\affiliation{US Naval Research Laboratory, Washington, D.C. 20375, USA}
\author{Kenji~Watanabe}
\author{Takashi~Taniguchi}
\affiliation{National Institute for Materials Science, 1-1 Namiki, Tsukuba 305-0044, Japan}
\author{Timur K. Kim}
\author{Cephise Cacho}
\affiliation{Diamond Light Source, Division of Science, Didcot, United Kingdom\\ $^{\dagger}$ These authors contributed equally\\ $^{\ast}$ Corresponding Authors: jkatoch@andrew.cmu.edu (J. K.) and ulstrup@phys.au.dk (S. U.)}
\author{Nicola~Lanata}
\author{Jill~A.~Miwa}
\author{Philip~Hofmann}
\affiliation{Department of Physics and Astronomy, Aarhus University, 8000 Aarhus C, Denmark}
\author{Jyoti Katoch$^{\ast}$}
\affiliation{Department of Physics, Carnegie Mellon University, Pittsburgh, Pennsylvania 15213, USA}
\author{S{\o}ren Ulstrup$^{\ast}$}
\affiliation{Department of Physics and Astronomy, Aarhus University, 8000 Aarhus C, Denmark}

\begin{abstract}
The possibility of triggering correlated phenomena by placing a  singularity of the density of states near the Fermi energy remains an intriguing avenue towards engineering the properties of quantum materials. Twisted bilayer graphene is a key material in this regard because the superlattice produced by the rotated graphene layers introduces a van Hove singularity and flat bands near the Fermi energy that cause the emergence of numerous correlated phases, including superconductivity. While the twist angle-dependence of these properties has been explored, direct demonstration of electrostatic control of the superlattice bands over a wide energy range has, so far, been critically missing. This work examines a functional twisted bilayer graphene device using \textit{in-operando} angle-resolved photoemission with a nano-focused light spot. A twist angle of 12.2$^{\circ}$ is selected such that the superlattice Brillouin zone is sufficiently large to enable identification of van Hove singularities and flat band segments in momentum space. The doping dependence of these features is extracted over an energy range of 0.4~eV, expanding the combinations of twist angle and doping where they can be placed at the Fermi energy and thereby induce new correlated electronic phases in twisted bilayer graphene.
\end{abstract}

\maketitle

In single-layer graphene a saddle-point in the dispersion of the $\pi$-band appears around the $\bar{\mathrm{M}}$-point of the Brillouin zone (BZ), leading to a van Hove singularity (vHs) in the density of states. Since this occurs far from the Fermi energy, $E_F$, extensive doping would be required to shift the vHs near $E_F$ \cite{McChesney:2010}, keeping this critical point out of reach of transport experiments with standard electrostatic gating techniques. Stacking two graphene layers with a controlled interlayer twist angle ($\theta$) elegantly solves this problem because the two rotated Dirac cones hybridize and form a new superlattice vHs with an energy that decreases as $\theta$ is reduced  \cite{Guohong:2010,Santos:2007,Bistritzer:2011,Morell:2010,Santos:2012}. This level of tunability not only makes the physics of the vHs accessible to electron-transport \cite{Cao:2016,Kim:2017,Cao:2018a,Cao:2018b,Yankowitz:2019,Xiaobo:2019} and -tunneling experiments \cite{Kerelsky:2019,Yonglong:2019,Jiang:2019,Choi:2019}, but also leads to the possibility of tuning the vHs resonantly with a desired optical excitation  \cite{Trambly:2012}, which has been explored in optical conductivity measurements of twisted bilayer graphene (twBLG) for a wide range of $\theta$ \cite{Tabert:2013,Havener:2014}.  

\begin{figure*} 
\begin{center}
\includegraphics[width=1\textwidth]{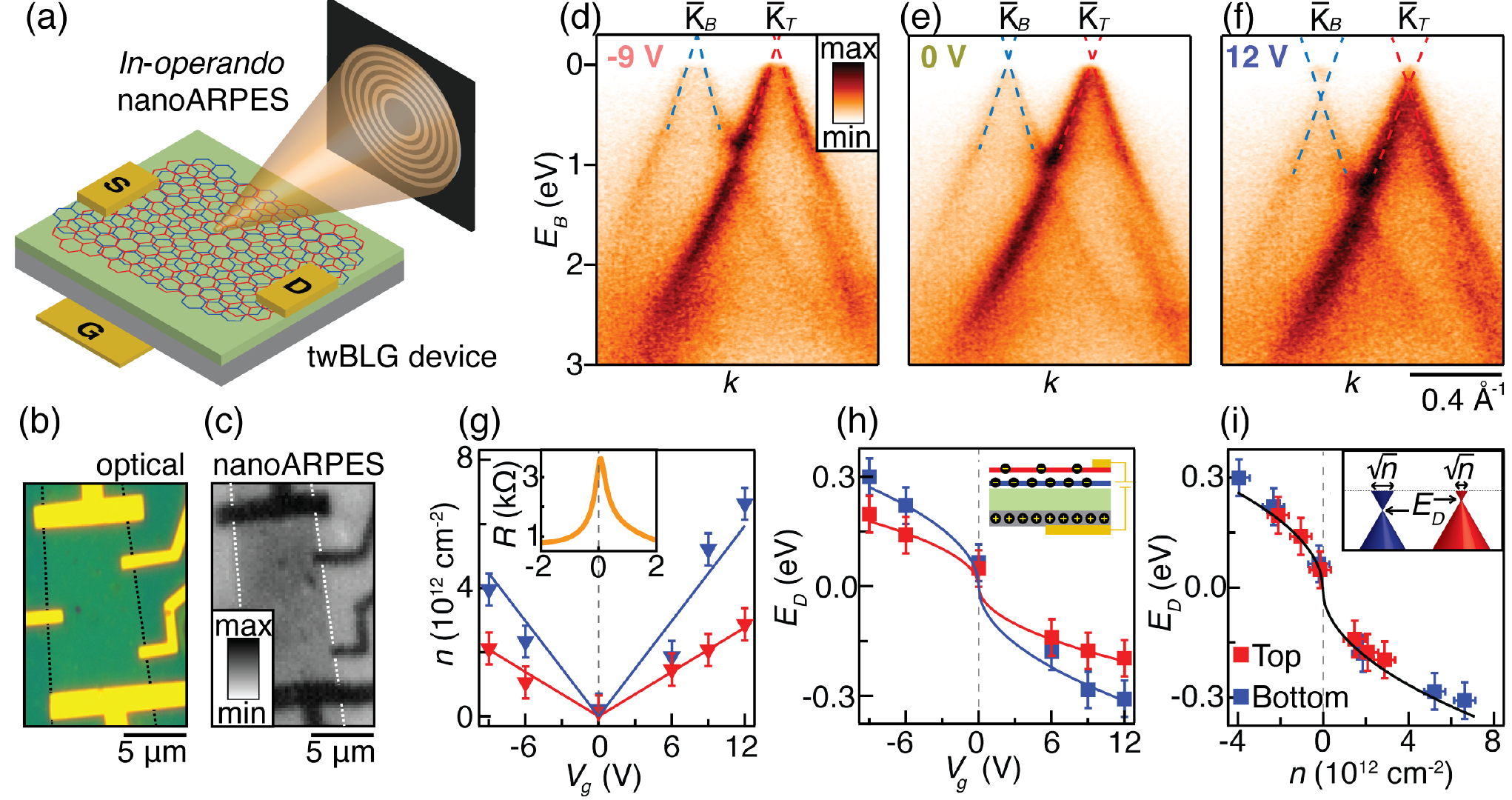}
\caption{Electrostatic tuning of twisted bilayer graphene Dirac cones viewed by nanoARPES. a) Sketch of our \textit{in-operando} nanoARPES experiment with top (red) and bottom (blue) graphene layers contacted by source (S) and drain (D) electrodes and stacked on hBN and on a graphite back-gate (G). A beam of photons is focused to a 690~nm spot using zone plate optics. b) - c) Functional region of device presented via (b) optical micrograph and (c) spatially-dependent ARPES intensity integrated over $E$ and $k$. The dotted lines demarcate the twBLG flake. d) - f) ARPES spectra of the twBLG Dirac cones measured at the given gate voltages. The dashed blue (red) lines represent linearly extrapolated peak positions determined from MDC fits of the bottom (top) layer Dirac cone around $\bar{\mathrm{K}}_B$ ($\bar{\mathrm{K}}_T$) as shown in Supplementary Figure S3. g) Gate voltage dependence of $n$ obtained from $k_F$ of each Dirac cone. The curves are fits to a linear dependence on gate voltage. The inset presents the resistance of the device measured in situ before exposure to photons. h) Dirac point energies $E_D$ determined from the linear extrapolation shown in (d)-(f) with fits to two separate $\sqrt{V_g}$-dependent functions (curves). The different slopes of the fitted curves in (g)-(h) result from the smaller amount of charge induced in the top layer (see capacitor model of our device in the inset). i) Demonstration of $\sqrt{n}$-dependence (black curve) of $E_D$ for bottom (top) graphene, as expected for non-interacting Dirac cones (see sketch of cones at different doping levels in the inset). The blue (red) markers represent the bottom (top) layer and vertical dashed lines mark the charge neutrality point in (g)-(i).}
\label{fig:1}
\end{center}
\end{figure*}

The energy- and momentum-dependent evolution of the interlayer hybridization between the Dirac cones of twisted graphene layers has been observed in angle-resolved photoemission spectroscopy (ARPES) experiments for large twists \cite{Ohta:2012,Peng:2017} and near the magic angle  \cite{flatbandArxiv1,flatbandArxiv2}, but the effect of gating in a functional device has not been previously explored. We achieve this by focusing a beam of 60~eV photons to a spot-size of 690~nm using a Fresnel zone plate, leading to angle-resolved photoemission with nanoscale spatial resolution (nanoARPES) from a twBLG flake supported on hexagonal boron nitride and back-gated by graphite as sketched in \textbf{Figure~\ref{fig:1}}a. Similar approaches have only recently been demonstrated to lead to electrostatically tunable bands in single-layer graphene \cite{Nguyen:2019,muzzio2020} and Bernal-stacked bilayer graphene \cite{Joucken:2019}. An optical micrograph of the device is compared to a map of the photoemission intensity in Figure~\ref{fig:1}b - Figure~\ref{fig:1}c, revealing the location of the twBLG flake between the top contacts used for grounding the flake. The data underlying the nanoARPES map is composed from a four-dimensional dataset containing the $(E,k,x,y)$-dependent photoemission intensity and the image merely represents a projection of the $E$- and $k$-integrated intensity onto real space. A more detailed analysis, shown in Supplementary Figure~S1 - Figure~S2, sharply outlines the individual components of the device and also demonstrates the separation of disordered from clean areas. Importantly, the results presented in Supplementary Figure S2 enable us to quantify a spatially varying small-scale rotational disorder of the graphene flakes, which were originally synthesized using chemical vapor deposition on copper foils. The interlayer twist angle of our stacked graphene is predominantly given by $\theta = 12.2^{\circ}$ but a variation of $\pm 1^{\circ}$ occurs across the device. Our nanoARPES approach enables us to single-out the domains with  $\theta = 12.2^{\circ}$ \cite{Joucken:2019n}, which is the focus of this study.

\begin{figure*} 
\begin{center}
\includegraphics[width=0.9\textwidth]{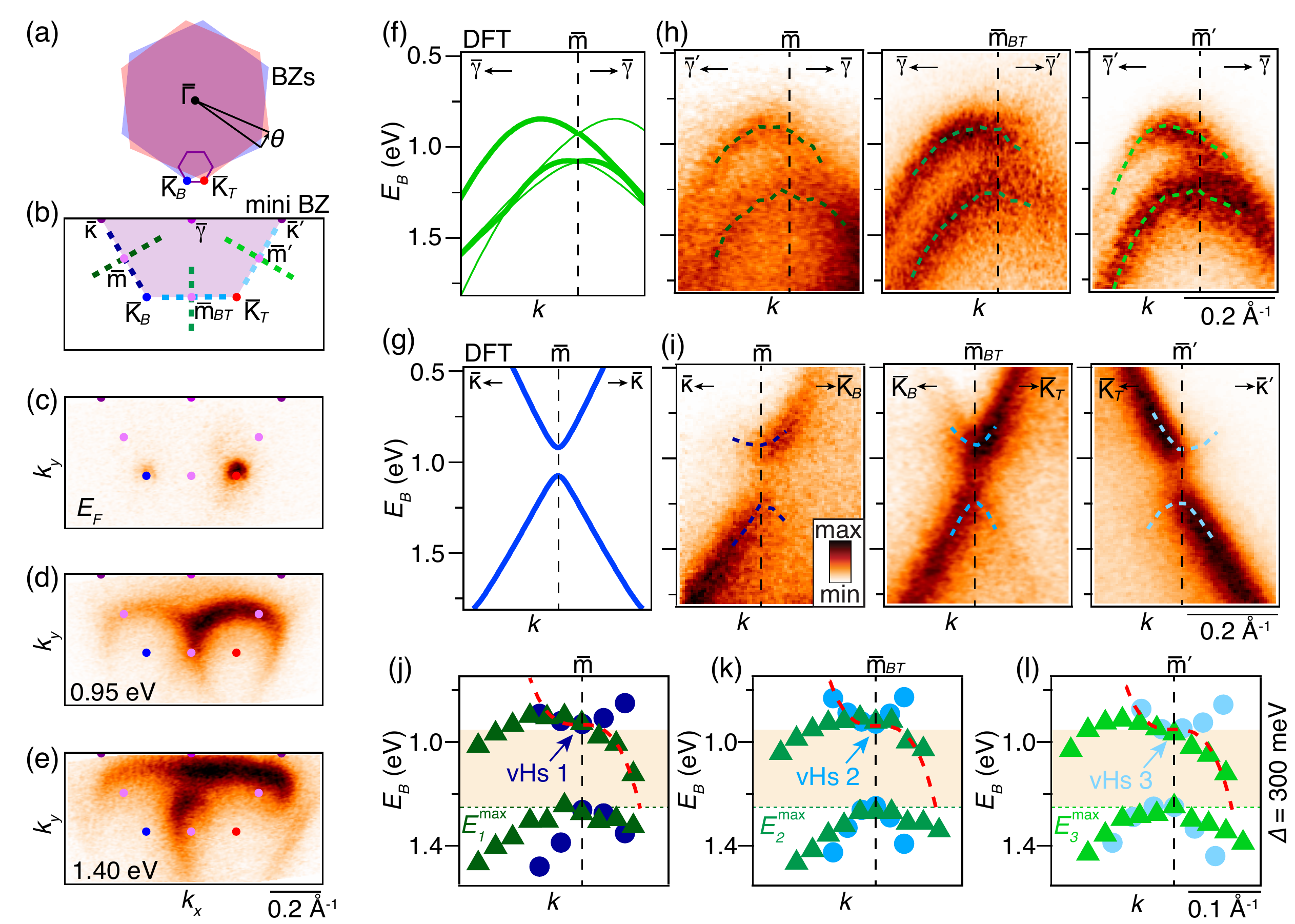}
\caption{Dispersion and van Hove singularities in the mini Brillouin zone. a) Main BZs with an interlayer rotation of $\theta =12.2^{\circ}$. The mini BZ is indicated along with the main Dirac points $\bar{\mathrm{K}}_B$ and $\bar{\mathrm{K}}_T$ for the bottom and top layer, respectively. b) Measured segment of the mini BZ with indication of high symmetry points. c) - e) Constant energy cuts at (c) the Fermi energy ($E_F$), (d) the energy of the superlattice vHs ($E_B = 0.95$~eV) and (e) at an energy below the intersection of the two Dirac cones ($E_B = 1.40$~eV). High symmetry points of the mini BZ are indicated by dots as shown in (b). f) - g) DFT bands mirrored around $\bar{\mathrm{m}}$ in the (f) $\bar{\mathrm{m}}$-$\bar{\gamma}$ and (g) $\bar{\mathrm{m}}$-$\bar{\kappa}$ directions. Thick lines indicate the bands identified in the ARPES spectra. h) - i) ARPES spectra representing the high symmetry directions indicated by (h) green and (i) blue dashed lines in (b). The overlaid dashed curves are results of EDC fits of the peak positions. j) - l) Combined results of the EDC fits around the three measured $\bar{\mathrm{m}}$-points in the orthogonal directions in (h) and (i), revealing a vHs and a mini gap given by $\Delta = 300 \pm 20$~meV (see shaded area) at each $\bar{\mathrm{m}}$-point. Each vHs is marked by an arrow and the corresponding saddle point is illustrated by a fit to a third order polynomial (red dashed curves). Horizontal dotted lines mark the maximum energy below the mini gap. Triangular and circular markers correspond to the fits in (h) and (i), respectively, and the size of the markers reflect the error bars associated with the determined band energies.}
\label{fig:2}
\end{center}
\end{figure*} 

A selection of ARPES spectra for the full range of available gate voltages is presented in Figure~\ref{fig:1}d - Figure~\ref{fig:1}f for a cut along a line connecting the bottom and top layer Dirac points $\bar{\mathrm{K}}_B$ and $\bar{\mathrm{K}}_T$, respectively. At our twist angle of $12.2^{\circ}$ the spectrum is characterized by two separated non-interacting graphene Dirac cones at low binding energies, before the two cones hybridize around a binding energy of 1~eV. The Dirac cone of the bottom graphene layer is less intense due to attenuation of the photoelectrons through the top layer. These observations are consistent with previous ARPES studies of twBLG flakes supported on copper films \cite{Peng:2017} and on silicon carbide \cite{Ohta:2012}. The linear part of each cone is analyzed using momentum distribution curve (MDC) cuts as demonstrated in Supplementary Figure S3, which leads to the extrapolated dashed linear branches seen in panels d-f. The energy of the Dirac point, $E_D$, and the magnitude of the Fermi wavevector, $k_F$, differ substantially in the hole- and electron-doped situations in panels d and f, respectively. Values of carrier density $n = k_F^2/\pi$ for top and bottom Dirac cones along with the resistance curve of our functional device are summarized in Figure \ref{fig:1}g. Both the $n(V_g)$ curves and the $E_D(V_g)$-relations seen in Figure \ref{fig:1}h display a change of slope between the two layers, reflecting the smaller amount of charge that is induced by the gate in the top layer (see illustration in the inset of Figure \ref{fig:1}h). Finally, by combining the results for $n$ and $E_D$ for each layer as shown in Figure \ref{fig:1}i we find that in both cases $E_D$ scales with doping as $\sqrt{n}$, confirming the isolated graphene-like behavior in each cone \cite{Nguyen:2019,muzzio2020}.

We now turn our focus towards the key features of the superlattice dispersion at zero gate voltage before we return to the electrostatic tunability of these features. \textbf{Figure \ref{fig:2}}a illustrates the two main graphene BZs rotated by $\theta = 12.2^{\circ}$ and the resulting mini BZ, which is constructed from the reciprocal moir\'e lattice vector given by $\vec{G}_m = \vec{G}_T - \vec{G}_B$, where $ \vec{G}_T$ ($ \vec{G}_B$) is the top (bottom) graphene reciprocal lattice vector. The length-scale of the mini BZ in terms of $|\vec{G}_m|$ can be related to the $k$-distance ($\Delta K$) between $\bar{\mathrm{K}}_B$ and $\bar{\mathrm{K}}_T$ as  $|\vec{G}_m| = \sqrt{3}\Delta K = 2\sqrt{3}|\bar{K}|\sin{(\theta/2)}$, where $|\bar{K}| \approx 1.7$~\AA$^{-1}$ is the distance to the main Dirac point from $\bar{\mathrm{\Gamma}}$. We obtain $|\vec{G}_m| = 0.63$~\AA$^{-1}$ for our twist angle, which is roughly a factor of 11 times larger than the mini BZ corresponding to the "magic angle" of 1.1$^{\circ}$ \cite{Bistritzer:2011}. In the latter case, the $k$-dependent linewidth of the features becomes comparable to the full size of the mini BZ \cite{flatbandArxiv1,flatbandArxiv2}, making a detailed analysis of features within a single mini BZ almost impossible. Here we exploit the larger mini BZ presented in Figure \ref{fig:2}b to track the evolution of the key features, which are the superlattice vHs and the mini gap $\Delta$, and are able to spectrally resolve them within the mini BZ. 

The constant energy cuts in Figure~\ref{fig:2}c - Figure~\ref{fig:2}e are presented on the same $k$-scale as the mini BZ in Figure \ref{fig:2}b, with the high symmetry points labeled by correspondingly colored dots. Two isolated Dirac points evolve from $E_F$ (panel c) and start touching at a binding energy of 0.95~eV at the point labeled $\bar{\mathrm{m}}_{BT}$ on the border of the mini BZ (see panel d). At a higher binding energy of 1.40 eV the cones fully hybridize, leading to knot-shaped contours around $\bar{\mathrm{m}}$, $\bar{\mathrm{m}}_{BT}$ and  $\bar{\mathrm{m}}^{\prime}$. These are signatures of hybridization between the main Dirac cones and the mini cones emerging from the $\bar{\kappa}$ and  $\bar{\kappa}^{\prime}$ points. The dispersion of the mini bands around $\bar{\mathrm{m}}$ is calculated in the two orthogonal high symmetry directions using Density Functional Theory (DFT) for a commensurate twBLG supercell with $\theta =$ 13.17$^{\circ}$ as shown in Figure~\ref{fig:2}f - Figure~\ref{fig:2}g. These band structures approximate the nanoARPES $(E,k)$-cuts around $\bar{\mathrm{m}}$, $\bar{\mathrm{m}}_{BT}$ and  $\bar{\mathrm{m}}^{\prime}$ shown in Figure~\ref{fig:2}h - Figure~\ref{fig:2}i. Intriguingly, we observe downwards curving bands along $\bar{\gamma}$-$\bar{\mathrm{m}}$ and upwards curving bands along $\bar{\kappa}$-$\bar{\mathrm{m}}$ around each of the measured $\bar{\mathrm{m}}$ points in the corresponding orthogonal directions. Note that the thin curves seen in the DFT results in Figure~\ref{fig:2}f are absent in the data in Figure~\ref{fig:2}h,  and an abrupt decrease of intensity towards the mini cones at $\bar{\kappa}$ and  $\bar{\kappa}^{\prime}$ occurs in the cuts shown in Figure~\ref{fig:2}i because of the incommensurate nature of the superlattice formed at our twist angle. The sign-change of the curvature, which is summarized via the results of energy distribution curve (EDC) fits in Figure~\ref{fig:2}j - Figure~\ref{fig:2}l (see also dashed lines in Figure~\ref{fig:2}h - Figure~\ref{fig:2}i and examples of EDC fits in Supplementary Figure S4) is evidence for a vHs at each mini BZ boundary at a binding energy of 0.95 eV. The corresponding saddle-point is further demonstrated by fits to third order polynomials shown via dashed red curves in Figure~\ref{fig:2}j - Figure~\ref{fig:2}l. A mini gap is formed at each $\bar{\mathrm{m}}$ point with a magnitude given by $\Delta = 300 \pm 20$~meV (see shaded area in Figure~\ref{fig:2}j - Figure~\ref{fig:2}l). This is substantially larger than the value of 150~meV predicted by the calculation and also than the $\approx$200~meV found in a similar twBLG stack on silicon carbide \cite{Ohta:2012}, possibly because a stronger interlayer interaction is established in our particular configuration.
 
 \begin{figure*} 
\begin{center}
\includegraphics[width=0.8\textwidth]{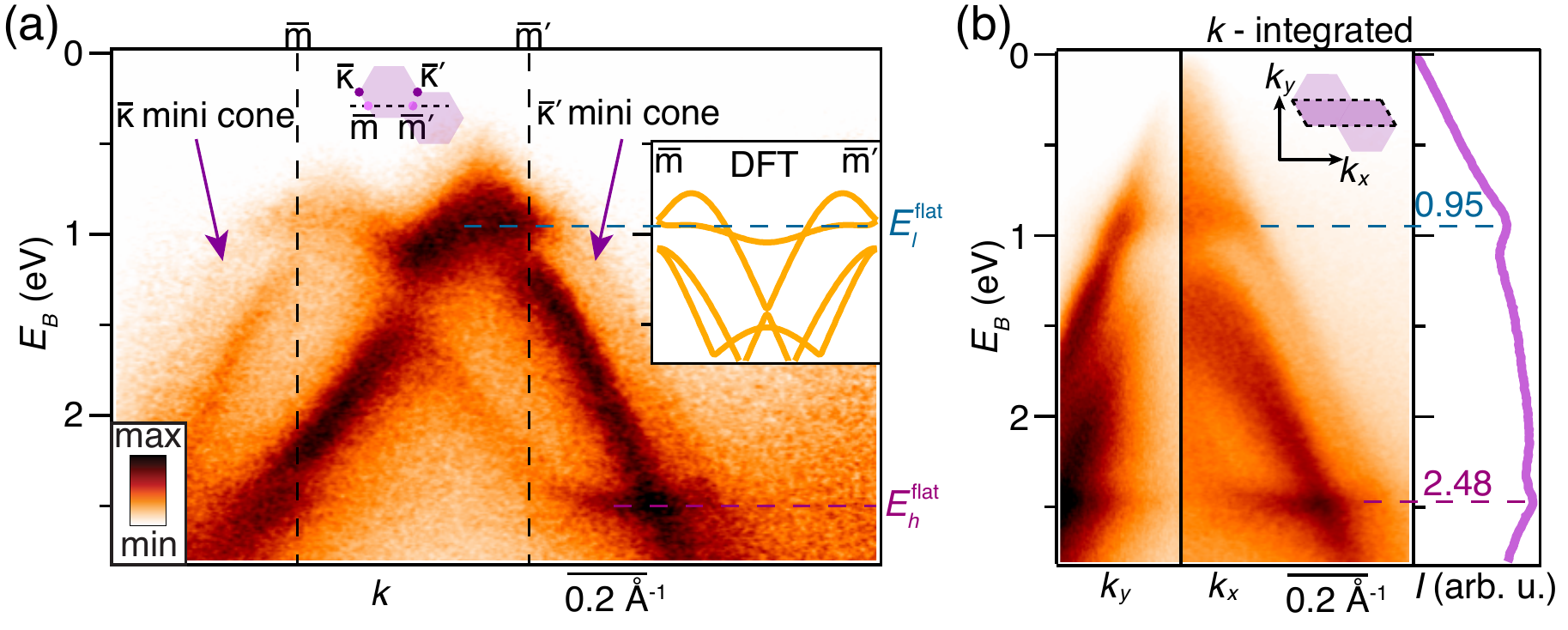}
\caption{Observation of flat band segments at multiple energies. a) ARPES intensity cut along a line that connects $\bar{\mathrm{m}}$ and  $\bar{\mathrm{m}}^{\prime}$ as shown in the sketch of neighboring mini BZs. The dispersion of the mini cones originating from $\bar{\kappa}$ and $\bar{\kappa}^{\prime}$ are indicated by arrows. The inset presents DFT bands for a twBLG superlattice with $\theta = 13.17^{\circ}$ plotted along the  $\bar{\mathrm{m}}$-$\bar{\mathrm{m}}^{\prime}$ direction. A flat band segment is observed around the superlattice vHs at a lower binding energy ($E^{\mathrm{flat}}_l$) and an additional segment occurs around the crossing between the the main cones at a higher binding energy ($E^{\mathrm{flat}}_h$). b) Photoemission intensity integrated over the $k_x$-direction (left panel), $k_y$-direction (middle panel) and $(k_x,k_y)$-plane (right panel) of a full mini BZ as sketched by the dark purple area enclosed by dashed lines in the mini BZ diagram in the middle panel. The colored dashed horizontal lines in (a)-(b) label the binding energies of the flat band segments $E^{\mathrm{flat}}_l$ and $E^{\mathrm{flat}}_h$ with the values given in units of eV in the right panel in (b).}
\label{fig:3}
\end{center}
\end{figure*}

\begin{figure*} 
\begin{center}
\includegraphics[width=1\textwidth]{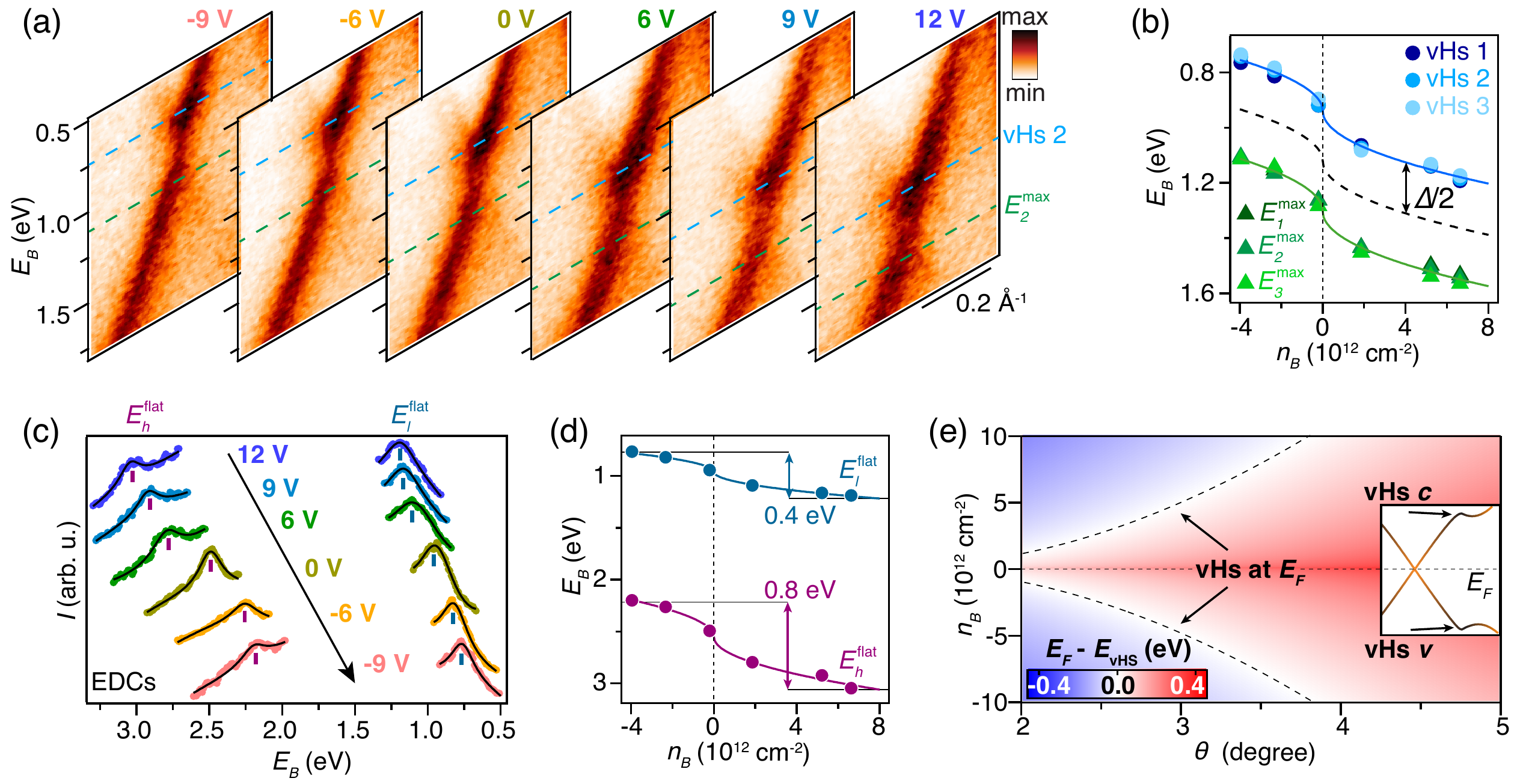}
\caption{Doping dependence of flat band segments and van Hove singularities. a) Snapshots of the crossing between the main cones at the given gate voltages. The dashed horizontal lines delineate the minigap $\Delta$, which separates the vHs 2 and the band maximum $E^{\mathrm{max}}_2$.  b) Binding energies of all vHs points and band maxima defined in Figure \ref{fig:2} as a function of carrier density ($n_B$) of the bottom graphene layer. The full curves are fits to a $\sqrt{n}$-dependence, and the dashed curve tracks the middle of the minigap (see double-headed arrow). c) EDCs (colored markers) of the flat band segments defined in Figure \ref{fig:3} for the given gate voltages. The ticks indicate the binding energy positions of the peaks fitted by a Lorentzian function on a linear background (black curves). d) Values of $E^{\mathrm{flat}}_l$ and $E^{\mathrm{flat}}_h$ corresponding to the ticks displayed in (c) with fits to a $\sqrt{n}$-dependence (curves). The complete energy range that is accessible by doping is indicated by a double-headed arrow for each flat band segment. e) Phase diagram describing the vHs binding energy with respect to the Fermi energy as a  function of twist angle and doping of the bottom graphene layer. The color scale refers to the vHs energies below (vHs $v$) and above (vHs $c$) the Dirac crossing (see arrows on the DFT bands plotted along $\bar{\kappa}$-$\bar{\mathrm{m}}$-$\bar{\gamma}$ in the inset) for hole-type and electron-type doping, respectively. The dashed curves demarcate the critical situation where a superlattice vHs is exactly at $E_F$.}
\label{fig:4}
\end{center}
\end{figure*}

The superlattice vHs is accompanied by a weakly dispersing segment (or flat band) \cite{Bistritzer:2011,Morell:2010}, which triggers strongly correlated phenomena when it is placed at $E_F$ because the local Coulomb energy in the moir\'e unit cell can exceed the bandwidth of these states \cite{Cao:2018a,Cao:2018b}. This flat segment of the dispersion is a result of hybridization and avoided-crossing effects between all of the mini cones around the vHs, which are also present in our data and reproduced by DFT calculations as demonstrated in the $\bar{\mathrm{m}}$-$\bar{\mathrm{m}}^{\prime}$ cut in \textbf{Figure \ref{fig:3}}a. In addition to this flat segment at the lower binding energy ($E^{\mathrm{flat}}_l$) of the vHs, situated at 0.95 eV, we also find a higher binding energy ($E^{\mathrm{flat}}_h$) flat segment at 2.48 eV. The latter occurs in a nonlinear part of the dispersion from the bottom graphene Dirac cone where it again crosses a branch of the top Dirac cone and is attributed to the dispersion around the $\bar{\mathrm{M}}$-point of the main BZ where the vHs of isolated graphene occurs \cite{McChesney:2010}. These flat segments stand out when we integrate the photoemission intensity in the two orthogonal directions of the mini BZ and appear as two well-defined peaks in the fully $k$-integrated intensity as shown in Figure~\ref{fig:3}b, which emphasizes their weak dispersion with $k$.

We now investigate how the vHs and flat segments shift with applied gate voltage. The series of snapshots along $\bar{\mathrm{K}}_T$-$\bar{\mathrm{K}}_B$ in \textbf{Figure \ref{fig:4}}a demonstrate that the intersection of the cones and the vHs move to lower binding energies with increasing hole-doping, i.e. with the application of increasingly negative gate voltage. Figure \ref{fig:4}b presents the complete doping dependence of the vHs and the mini gap determined from an EDC analysis at each mini BZ boundary (see Supplementary Figure S4). The positions of the features follow a $\sqrt{n}$-dependence, while the magnitude of the gap remains fixed. A similar dependence is observed for the flat band segments extracted from the EDC analysis presented in Figure \ref{fig:4}c and summarized in Figure \ref{fig:4}d. Interestingly, the overall energy range that can be tuned by doping is doubled for the higher binding energy component. The wider energy range is attributed to the combination of different charge induced in the two graphene layers and the shift taking place in a highly non-linear part of the bottom Dirac cone.  

By combining the extracted $\sqrt{n}$-dependence of the vHs at our twist angle with its known dependence on twist angle we are able to construct a ($\theta$,$n$)-phase diagram that describes the position of the vHs energy measured from $E_F$ as shown in Figure \ref{fig:4}e, extending our results to a wide range of twBLG systems. The $\theta$-dependence is given by $E^{\pm}_{vHs}(\theta)=  \pm\hbar v^{\star}(\theta)\Delta K(\theta)$, where the negative (positive) sign selects the singularity below (above) the Dirac crossing, which we label vHs $v$ (vHs $c$) as shown in the DFT calculation in the inset of Figure \ref{fig:4}e. Furthermore, $v^{\star}/v_F = (1-3\alpha^2)/(1+6\alpha^2)$ describes the modification of the Fermi velocity $v_F$ caused by the hybridization between the Dirac cones through the parameter $\alpha = w/\hbar v_F \Delta K$ where we used $w = 110$~meV \cite{Santos:2007,Bistritzer:2011,Santos:2012,Trambly:2012}. This leads to the diagram presented in Figure \ref{fig:4}e where we display the range of angles where the vHs $v$ and vHs $c$ energies can be placed at $E_F$ (see dashed curves) using the order of magnitude of electrostatic doping that we have demonstrated in this work. Note that below $\theta = 2^{\circ}$ the velocity of the Dirac particles develops a more complex behavior that establishes the magic angle series \cite{Bistritzer:2011,Guohong:2010}. 

In conclusion, we have used nanoARPES to identify the vHs locations and flat band segments in the mini band dispersion of twBLG at a twist angle of $\theta = 12.2^{\circ}$. Electrostatic doping of the two graphene layers has enabled us to track the $(E,k)$-dependence of these features, expanding the range of twist angles and doping where they can be placed at the Fermi energy. These results lead to the tantalizing prospects of triggering novel interactions for larger twist angles and thereby larger mini BZs that can be fully explored in $E$- and $k$-resolving experiments combined with electron transport. Most importantly, our work establishes a route forward for accessing emerging phenomena associated with vHs physics for a wider range of superlattice heterostructures.

\section{Methods}

\textit{Graphene growth on copper foil.} Graphene was grown on a 25~$\mu$m thick copper foil using low-pressure chemical vapor deposition (CVD) with H$_2$ and CH$_4$ gases \cite{Zhu:2011g}.  Prior to growth, the copper foil was electrochemically polished to remove oxides and improve surface quality.  The foil was folded into a pouch by folding it in half and then folding three edges to create an enclosure \cite{Xuesong:2011}.  The packet was then placed into a quartz tube which was brought to a base pressure of 2 mTorr.  Next, the copper was heated to 1030~$^\circ$C and H$_2$/CH$_4$ gas was circulated in the chamber for 1.5~hours with a total pressure of $\approx$60~mTorr.  The substrate was quenched through removal from the hot zone under vacuum.\\

\textit{Heterostructure Preparation.} Bulk hBN and graphite crystals were mechanically exfoliated using the scotch tape method on two separate O$_2$ plasma treated SiO$_2$(300~nm)/Si substrates. Exfoliated graphite and hBN flakes of thickness ~10~nm and ~30~nm, respectively, were selected optically. The hBN/graphite stack was made with a custom-built transfer tool by using a thin PC film on top of PDMS to first pick up the hBN flake from the substrate and then subsequently use the hBN to pick up the graphite flake \cite{katoch2018}. This hBN/graphite stack was then dropped onto a SiO$_2$/Si substrate with pre-written bond pads and large electrodes. The stack was annealed at 350 $^{\circ}$C for 15 minutes in high vacuum in order to remove polymer residue on the surface of the hBN. The CVD graphene sheets were obtained by cutting a section of the copper foil (with graphene grown on both sides).  One side of the Cu foil with graphene was coated in PMMA which acts as a protective layer and the other side of the foil was exposed to O$_2$ plasma in a reactive ion etcher to remove graphene.  The copper foil was then cut into small pieces ($\approx$6 mm$^2$) and floated on a copper wet etchant where the exposed copper was in contact with the etchant \cite{Patra:2012}.  After the copper was fully etched, graphene with PMMA on top was transferred onto the hBN/graphite stack, covering the majority of the SiO$_2$/Si wafer. The heterostructure was baked at 180 $^{\circ}$C on a hot plate for 5~minutes followed by removal of PMMA by dipping in acetone.  Similarly, another CVD graphene layer was placed on the heterostructure and cleaned using the same procedure to complete the twBLG/hBN/graphite heterostructure. Stacking two large-area CVD graphene layers has been demonstrated to lead to hybridization of the layers, despite the chemical residues that exist \cite{Robinson:2013a,Beechem:2014}.\\

\textit{Device Fabrication.} Electron beam lithography, using a MMA/PMMA bilayer resist, was employed to define an etch pattern to remove the graphene everywhere except for a stripe on the hBN. The exposed graphene was then etched using O$_2$ plasma in a reactive ion etcher.  The resulting stack was annealed in high vacuum at 350 $^{\circ}$C for 15~minutes.  To ensure electrical contact to the graphite back gate, holes were bored into the hBN to expose the underlying graphite.  A final electron beam lithography step was performed to define electrodes. A 5~nm film of Cr and 110~nm film of Au were deposited using electron beam deposition. The wafer containing the device was finally placed in a CSB00815 chip package and wire-bonded.\\

\textit{nanoARPES.} The measurements presented here were performed at the nanoARPES branch of the I05 beamline of Diamond Light Source. The chip carrier with device was annealed in ultra-high vacuum at 150 $^{\circ}$C for 90 minutes in the nanoARPES preparation system before exposure to the synchrotron beam. 

We used a synchrotron beam with a photon energy of 60 eV that was focused via a Fresnel zone plate to a spot size of 690~nm as demonstrated in Supplementary Figure S5. Photoemission spectra were obtained using a Scienta Omicron DA30 hemispherical analyser, with angular and energy resolution set to 0.2$^{\circ}$ and 30~meV, respectively. The scans of ($E$,$k_x$,$k_y$)-dependent photoemission intensity were obtained using the deflector mode of the DA30 analyser. The four-dimensional data sets containing the ($E,k,x,y$)-dependent intensity were collected by rastering the sample position relative to the beam with piezoelectric stages and measuring an ($E,k$)-dependent spectrum at each position. The sample was aligned such that each $(E,k)$-snapshot was collected in a direction of the BZ that is perpendicular to the $\bar{\mathrm{\Gamma}}$ - $\bar{\mathrm{K}}_{T}$ direction. The sample was kept at a temperature of 70~K during measurements.

Electrical doping of the device was achieved by applying a voltage to the graphite back gate while keeping the graphene flake at ground. Maximum and minimum gate voltages were determined by the onset of a leakage current through the insulating hBN gate dielectric. A 1 mV voltage was applied between source and drain contacts for each gate voltage in order to obtain the resistance curve presented in Figure~\ref{fig:1}g.\\ 

\textit{Density Functional Theory calculations.} All the DFT calculations were performed using the Vienna $Ab$ $initio$ Simulation Package (VASP) code \cite{kresse1996efficiency,kresse1999ultrasoft}. The exchange-correlation potentials were described through the Perdew-Burke-Ernzerhof (PBE) functional within the generalized gradient approximation (GGA) formalism \cite{perdew1996generalized}. A plane wave basis set was used with a cutoff energy of 400 eV on a $8\times8$ Monkhorst-Pack \cite{monkhorst1976special} $k$-point mesh. Graphene lattice constant and interlayer distances have been adapted from Ohta $et$ $al$ \cite{Ohta:2012}. A vacuum region of 16~\AA~along the $z$ direction (orthogonal to the layer plane) was used to separate the two heterostructures in order to minimize the interaction between the periodic repetitions of the cell, and the zero damping DFT-D2 method of Grimme \cite{grimme2006semiempirical} is used to account for the long range vdW interaction between graphene monolayers. For the modeled twist angle of 13.17$^{\circ}$, the commensurate moir\'e superlattice consists of 76 atoms and the moir\'e BZ is the same as the mini BZ (defined based on the difference of respective reciprocal vectors of the two graphene layers) demonstrated throughout this work.

\section{Supporting Information}
Supporting Information is available with the peer reviewed version.

\section{Acknowledgement}
We thank Diamond Light Source for access to Beamline I05 (Proposal No. SI24072) that contributed to the results presented here. S. U. acknowledges financial support from VILLUM FONDEN under the Young Investigator Program (Grant No. 15375). J. K. and R. M. acknowledge the financial support from U.S. Department of Energy, Office of Science, Office of Basic Energy Sciences, under Award Number DE-SC0020323. This work was supported by VILLUM FONDEN via the Centre of Excellence for Dirac Materials (Grant No. 11744) and the Independent Research Fund Denmark under the Sapere Aude program (Grant Nos. 9064-00057B and 6108-00409). K.W. and T.T. acknowledge support from the EMEXT Element Strategy Initiative to Form Core Research Center, Grant Number JPMXP0112101001 and the CREST(JPMJCR15F3), JST. S. P. acknowledges supports from Spanish MINECO for the computational resources provided through Grant FIS2016-80434-P.

\end{document}